\documentclass[10pt, twocolumn, nofootinbib]{revtex4-1}
\usepackage{amsmath,amssymb,amsthm}

\begin{document}

\title{Weak Cosmic Censorship, Superradiance and Quantum Particle Creation}
\author{\.Ibrahim Semiz}\email{semizibr@boun.edu.tr}
\author{Koray D\"{u}zta\c{s}}\email{koray.duztas@boun.edu.tr}
\affiliation{Bo\u{g}azi\c{c}i University, Department of Physics \\ Bebek 34342, \.Istanbul, Turkey}

\begin{abstract}
Since 1970's, gedanken experiments have been devised to challenge the weak cosmic censorship conjecture (WCCC), which is the expectation that spacetime singularities will be hidden from faraway observers by event horizons so that classical predictability in a spacetime is preserved. These experiments involve the interaction of an extremal or a slightly sub-extremal black hole with a test particle or field, attempting to destroy the horizon, i.e. to create a so-called naked singularity. They usually conclude that WCCC cannot be violated starting from an extremal black hole, but may be violated starting from a slightly sub-extremal one, if backreaction and self-force effects are neglected. Some other works also analyze these effects.

Starting 2007, a string of papers argue if WCCC can be violated by classically forbidden interactions occuring via the quantum nature of the particles associated with the fields; and where backrection and/or superradiance are pointed out as effects working in the direction of preserving the WCCC. We correct/modify a backrection argument, and furthermore point out that superradiance does not prevent {\em single particles} from being captured by the black hole; even if this capture would lead to WCCC violation. 

Then we consider the spontaneous emission (which we call the Zel'dovich-Unruh effect) of particles by the black hole, and find that at least for scalars, it can be understood without second quantization. It also completely invalidates the mentioned single- or few-particle thought experiments. However, the conclusions of our previous work on (at least) scalar fields interacting with black holes, i.e. that WCCC may be violated starting from slightly subextremal black holes, remains valid in this (semi)classical framework. 
\end{abstract}

\pacs{}

\maketitle

\section{Introduction: Challenging the Weak Cosmic Censorship Conjecture}

The deterministic nature of a spacetime in general relativity relies on the validity of Cosmic Censorship Conjecture which in its weak form (WCCC) states that gravitational collapse of a body always ends up in a black hole rather than a naked singularity~\cite{penrose.orig.ccc}, i.e. ``naked singularities" cannot evolve starting from nonsingular initial data. Conjecturing singularities to be hidden behind event horizons without any access to distant observers, enables the specification of a well defined initial value problem. The formation of singularities is inevitable once certain conditions become satisfied~\cite{singthm};  and if one were naked, it would prevent the existence of a Cauchy surface for the spacetime.

Not much progress has been made towards a concrete proof of the CCC, weak or strong~\cite{strongCCC}. Therefore one way to test the validity of the conjecture has been to challenge its seemingly weak spots by constructing gedanken experiments. In these experiments one envisages a black hole absorbing some particles or fields coming from infinity. The no-hair theorem \cite{mazur} in classical general relativity which states that stationary, asymptotically flat spacetimes are uniquely parametrized by three\footnote{If magnetic charge ($Q_{m}$) exist in nature, that will also be one of the parameters, bringing the total to four. It will enter  the metric via $Q^{2}=Q_{e}^{2}+Q_{m}^{2}$.} parameters (Mass $M$, charge $Q$, and angular momentum per unit mass $a$), guarantees that once the particles/fields are absorbed/reflected, the spacetime will settle to another spacetime with new parameters $M'$, $Q'$, and $a'$. The existence of the event horizon, which discriminates black holes and naked singularities, depends on an inequality involving these  parameters
\begin{equation}
M^{2} \geq Q^{2} + a^{2}. \label {criterion}
\end{equation}
in appropriate units~\cite{mtw}. In other words, a spacetime described by the Kerr-Newman metric corresponds to a black hole, if (\ref{criterion}) is satisfied; but to a naked singularity if it is violated; in the borderline case, i.e. when  (\ref{criterion}) is saturated, the spacetime is said to describe an extremal, or a critical black hole. The thought experiments are constructed to check if we can push the initially nonsingular spacetime satisfying (\ref{criterion}) beyond the extremal limit, so that the final spacetime violates (\ref{criterion}), and describes a naked singularity. To simplify the equation of motion of the incoming particle/field, they are taken as test particles/fields. Then, the changes they cause in the Kerr-Newman parameters are infinitesimal, therefore one should start the thought experiment from conditions infinitesimally close to where we would like to push the system, i.e. the extremal black hole. The first thought experiment in this vein was constructed by Wald in 1974~\cite{wald74}. He showed that particles with enough charge and/or angular momentum to overcharge/overspin a black hole either miss, or are repelled by, the black hole. This result was generalised to the case of dyonic black holes for spinless test particles \cite{hiscock} and scalar test fields \cite{dkn,toth}. These results suggest that the WCCC cannot be violated {\em quasi-statically}.

In 1999, Hubeny~\cite{hubeny} had the idea of starting from a nearly extremal Reissner-Nordstr\"{o}m, (i.e. nonrotating) black hole instead of an extremal one, and argued that the  black hole can be overcharged by using tailored test particles. Pursuing this avenue of thought requires careful gauging of the incoming particles/fields to be infinitesimal enough to be amenable to calculation, but not so infinitesimal as to rule out WCCC violation by the Wald-Hiscock-Semiz line of arguments; and it is hard to say that agreed-upon criteria for this fine-tuning exist. Nevertheless, this argument was adapted to Kerr (i.e. uncharged, rotating) \cite{Jacobson-Sot}, and extended to Kerr-Newman (i.e. fully general) black holes~\cite{saa_santarelli}. Later work \cite{barrauseEtal1,barrauseEtal2,isoyamaEtal,zimmermanEtal} considered backreaction, radiative and self-force effects neglected above, and concluded that these effects may prevent the particle from being captured. We analyzed the same question for test {\em fields} instead of particles for the Kerr black hole, and found similar results for fields of integer spin~\cite{overspin}, and somewhat more drastic ones for spin-1/2~\cite{koraydirac} (more on this in Sect.~\ref{ss:fieldsCCC} below). There are also works challenging WCCC with spherical shells~\cite{hod3} and claiming WCCC violation even for extremal black holes and test particles, due to higher order terms~\cite{gao_zhang}. 

%%%%%%%%%%%%%%%%%%%%%  SECTION II
\section{The Quantum connection} \label{Sect:quant_conn} 
%%%%%%%%%%%%%%%%%%%%

The generally accepted result of above works is that at least extreme black holes cannot be turned into naked singularities by absorbing classical test particles or fields. The particles are prevented from entering the black hole by a combination of the centrifugal and electric potentials and magnetic effects.

However, it is well-known that matter has ultimately a quantum nature; and quantum mechanics allows, for example, particles to go where they are classically forbidden to; a phenomenon known as tunneling. Therefore it is natural to wonder if the same phenomenon could allow test particles to tunnel through the barriers of the above-mentioned potentials. These considerations motivated Matsas \& da Silva~\cite{matsas_daSilva} to construct a gedanken experiment where neutral massless scalar quantum particles tunnel into a nearly extreme Reissner-Nordstr\"{o}m (RN) black hole. Obviously, to violate eq.(\ref{criterion}), the angular momentum/energy ratio should be large for the incoming particle. The authors investigate the Klein-Gordon equation on the RN metric to calculate the "absorption probability" in the low frequency (therefore low particle energy) limit, find that it is nonzero  for all nonzero frequencies, hence conclude that eq.(\ref{criterion}) can be violated if a low-energy particle with
\begin{equation}
l(l+1) > M^{2}(M^{2}-Q^{2})   \label{MdS_viol_crit}
\end{equation}
is absorbed by (tunnels into) the black hole where $l$ is the angular momentum quantum number of the particle, and $M$ and $Q$ are the initial mass and charge of the black hole.

Hod in \cite{hod1} considers backreaction for the same process. He points out that the black hole will acquire angular momentum through the process, and argues that precisely those frequencies that would lead to a violation of WCCC are prevented from entering the black hole by the phenomenon of superradiance~\cite{zeldovich,zeldovich2}: For 
\begin{equation}
\omega < \omega_{\rm sl} = m\Omega = m \frac{a}{r_{+}^{2}+a^{2}} = \frac{ma}{2Mr_+} ,  \label{sr_limit}
\end{equation}
the "reflection coefficient" is greater than unity. Here, $\Omega$ is the angular velocity of the event horizon, and \mbox{$r_{+} = M + \sqrt{M^{2}+a^{2}}$} its radius. The author calculates the value of $M^{2}-Q^{2}-a^{2}$ [cf. eq.(\ref{criterion})] after the process, with the minimum energy (i.e. frequency) given by eq.(\ref{sr_limit}), where the $a$ is the value acquired during the process. He finds that eq.(\ref{criterion}), therefore the WCCC cannot be violated\footnote{Hod in this work also reconsiders the Hubeny thought experiment, and argues that the gravitationally-induced self-force on the particles prevents the capture of those particles that would have led to violation of the WCCC.}.

On the other hand, fermions do not exhibit superradiance~\cite{no_ferm_SR}. This fact led Richartz \& Saa~\cite{ri_saa_1} to suggest replacing the scalar particles of the above thought experiments by fermions (still massles, hence they can be thought of as neutrinos), and trying to overspin a nearly extremal Kerr black hole. They consider a black hole one unit short (in Planck units, that is, in units determined by angular momentum quantization) from extremality, show that absorption of a low-energy particle with angular quantum numbers $l=m=3/2$ can lead to violation of WCCC, calculate the frequency interval needed for the violation; and since the "transmission coefficient" is positive for positive frequencies (which is another way of saying that there is no superradiance), they conclude that the particle will be absorbed, i.e. WCCC will be violated and the horizon destroyed. Alternatively they suggest that using a minimally charged black hole (still rotating almost extremely) will allow the use of an $l=m=1/2$ particle.  They also argue that using a large enough black hole, the backreaction issues mentioned above can be avoided. 

Hod in \cite{hod2} argues that vacuum polarization in the ergosphere of a rotating black hole will give rise to spontaneous emission of neutrinos; and this emission will both keep pushing the black hole away from extremality, and also suppress the absorption of incoming neutrinos due to the Pauli exclusion principle; thereby protecting WCCC.

Parallel to the discussion about fermions, Matsas et al. in~\cite{matsasEtal} counter the backreaction argument of Hod~\cite{hod1} by pointing out that in the thought experiment of~\cite{matsas_daSilva}, the total angular momentum transferred to (the originally nonrotating) black hole will be determined {\em not by $m$, but by $l$};  hence by preparing the incoming particle in an $m=0$ state,  the backreaction can be avoided and WCCC violated\footnote{They also conjecture that a naked singularity revealed or allowed by quantum effects might decay into elementary particles, whose entropy could preserve the generalized second law of thermodynamics.}. 

A more recent work by Richartz \& Saa~\cite{ri_saa_2} repeats the tunneling argument for almost extremal Reissner-Nordstr\"{o}m black holes and charged quantum particles, both scalars and spin-$1/2$ fermions; reaching similar conclusions. They also argue that the particles suitable for WCCC violation can be captured since the probabilities for such capture, as calculated thermodynamically~\cite{thermo_capture_prob} or by Quantum Field Theory (QFT)~\cite{qft_capture_prob} allow them to.

We believe that the disagreements in these works  partly stem from applying concepts relevant only for fields to cases involving particles. These concepts are the transmission and reflection coefficients, and superradiance. We turn to the discussion of these concepts, and their contexts of relevance in the next section.

%%%%%%%%%%%%%%%%%%%%%  SECTION III
\section{Classical vs. Quantum, Fields vs. Particles on Black Hole Spacetimes}
%%%%%%%%%%%%%%%%%%%%% 

Let us define/clarify our terms: By a classical field, we mean an entity obeying one of the well-known equations, it and physical quantities associated with it --such as energy and momentum densities-- being distributed over the spacetime in a continious fashion. When we say that we use a quantum field, however, we mean that {\em particles} are actually moving in spacetime, their behavior being ``guided by'' the field in a probabilistic way; allowing for calculation of {\em expectation values} of physical quantities of interest. This formalism is sometimes called first quantization. By the nature of the probabilistic description, these expectation values become better predictors of measurement results with increasing number of particles. When we deal with small number of particles, however, we will talk about using quantum particles. In this case, we should be thinking in terms of the second quantization formalism; that is, Fock spaces populated by states labeled by particle numbers and acted on by creation and annihiliation operators. In either case we consider {\em test} fields or particles, i.e. the effect of the fields/particles on the geometry is negligible, or can be estimated as a first order perturbation.

Of course, classical and quantum fields obey the same equations. On black hole spacetimes, some of these equations have been shown to be separable in Boyer-Lindquist coordinates~\cite{mtw}, which facilitates treatments of these fields. The most important such work is Teukolsky's separation~\cite{teuk2} of massless scalar, spin-$1/2$, spin-1 and spin-2 fields on the Kerr spacetime;
%\footnote{However, at least for the electromagnetic field (spin-1), the separation is valid for the Kerr-Newman metric as well, since the latter is obtained from Kerr by replacing $\Delta=r^2-2Mr+a^2$ with $\Delta=r^2-2Mr+a^2+Q^{2}$ (where $Q^{2}$ can be $Q_{e}^{2}+Q_{m}^{2}$ if the black hole has electric charge  $Q_{e}$ and magnetic charge $Q_{m}$). $Q$ does not appear outside $\Delta$, including the derivatives of $\Delta$, hence the  derivatives of the metric, which might contribute to the field equations. On the other hand, the electromagnetic field carries no charge, hence the free field will not couple to the field of the black hole, so no $Q$ terms will come from there either.};
the most general (complex, massive) free scalar~\cite{dkn-kg-sep} and spin-1/2~\cite{dkn-dirac-sep} fields have also been separated on the most general classical (dyonic Kerr-Newman) black hole. In the scalar or massless cases, the angular equations are Sturm-Liouville eigenvalue problems, therefore the eigenfunctions are both complete, so that the totality of the eigenmodes can represent the general solutions, and orthogonal, so that the modes can do so independently, one by one.

%%%%%%%%%%%%%%%%%%%%% 
\subsection{Classical fields, superradiance and WCCC} \label{ss:fieldsCCC}

To set the stage, and establish some notation; we briefly review the simplest field, but in its full generality: A mode of the massive complex scalar field can be written as $\Psi (r,\theta,\phi,t)=R(r) \Theta(\theta) e^{im\phi}e^{-i\omega t}$ where each factor satisfies its own equation.

As discussed in some of the references above, the $\Theta(\theta)$ functions can be orthonormalized, and the radial equation is transformed into
\begin{equation}
\frac{d^{2}}{dr_{*}^{2}} U(r_{*}) + V_{\rm eff}(r_{*}) U(r_{*}) = 0
\label{eq:radial.u}
\end{equation}
where $r_{*}$ is the well-known `tortoise' coordinate,  \mbox{$U(r_{*})=\sqrt{r^{2}+a^{2}}R(r)$}, and $V_{\rm eff}(r_{*})$ a complicated rational function of $r$ (see e.g. \cite{dkn} for its form) that reduces to two constants at the two ends, that is, to $\bar{\omega}^{2}$ near the horizon ($r \rightarrow r_{+}$, $r_{*} \rightarrow -\infty$), and to $(\omega^{2}-\mu^{2})$ near infinity (as $r_{*} \rightarrow r \rightarrow \infty$). Here we have
\begin{equation}
\bar{\omega} = \omega + 
\frac{eQ_{e}r_{+} - am}
             {r_{+}^{2}+a^{2}} 
\end{equation}
where $e$ is the charge and $\mu$ the mass of the field quantum.  
 
The boundary condition that nothing should come out of the (future) horizon of the black hole is usually adopted, restricting us to the solution
\begin{equation}
\lim_{r \rightarrow r_{+}} U_{l m}(\omega;r_{*})
= B_{l m}(\omega) e^{-i \bar{\omega}r_{*}} \label{horizon_solution}
\end{equation}
and for the same mode,
\begin{equation}
\lim_{r \rightarrow \infty} U_{l m}(\omega;r_{*})
                          = e^{-i k r_{*}} + A_{l m}(\omega) e^{i k r_{*}}
                          	\label{solution_at_infty}
\end{equation}
where $k^2 = \omega^{2} - \mu^{2} > 0$ for scattering states, and the solution has been normalized such that it corresponds to a wave of unit amplitude coming in from infinity, being transmitted into the black hole with amplitude $B$ and reflected back to infinity with amplitude $A$. The two amplitudes are related by the constancy of the Wronskian of the solution and its complex conjugate, since eq.(\ref{eq:radial.u}) is real, giving
\begin{equation}
\lim_{r \rightarrow r_{+}} W = \lim_{r \rightarrow \infty} W \Longrightarrow \bar{\omega} B B^{*}
 = k (1-A A^{*})  \label{eq:wronski}
\end{equation}
where the labels $\omega l m$ are implied for $A$ and $B$. If $\bar{\omega}<0$, $A A^{*}$ will be bigger than unity, meaning that a stronger wave will come back than sent into the black hole, i.e. superradiance.

The phenomenon of superradiance plays a role in gedanken experiments attempting to violate WCCC using bosonic fields impinging on a slightly subextremal black hole~\cite{overspin}. Violation is not possible at frequencies higher than $\omega_{0} = m/2M$ ($m$ being the azimuthal wave number), and at lower frequencies superradiance works against it. Hence, violation can be achieved in a narrow range $\omega_{\rm sl} < \omega < \omega_{1}$, where $\omega_{1} < \omega_{0}$, the range shrinking to zero as extremality is approached.

On the other hand, for {\em fermion fields}, superradiance does not occur~\cite{no_ferm_SR}, which was the motivation for~\cite{ri_saa_1}. In this case, the WCCC-violating range is not bounded   from below any more by superradiance~\cite{koraydirac}. Hence, unlike the bosonic case, this range does {\em not} shrink to zero as extremality is approached, therefore, contrary to implication of~\cite{ri_saa_1} (and its follow-up, \cite{ri_saa_2}), the violation will also work for extremal black holes. As far as we know, this is the only context where a thought experiment can result in destruction of an extremal black hole {\em without fine-tuning}. Claims of destruction of extremal black holes with finely tuned particles/fields, e.g.~\cite{gao_zhang}, may be challenged by backreaction or self-force effects, for example; but these thought experiments seem to be more robust (unless classical or first-quantized fermion fields are meaningless~\cite{QFT-Topo}). 

%%%%%%%%%%%%%%%%%%%%% 
 
\subsection{The meaning of transmission and reflection coefficients}

Because eq.(\ref{eq:radial.u}) looks like a \mbox{one-dimensional} Schr\"{o}dinger scattering problem with two (different)
constant potentials at two ends and a complicated potential well in the middle, one can define so-called transmission and reflection coefficients in analogy with that nonrelativistic problem. Since $k$ and $\bar{\omega}$ are the "wave numbers" in the $r_{*}$ coordinate, $AA^{*}$ would be the reflection coefficient, and $\bar{\omega} B B^{*}/k$ would be the transmission coefficient\footnote{Note that all of \cite{matsas_daSilva,hod1,ri_saa_1,hod2,matsasEtal,ri_saa_2} consider massless cases, for which $k$ becomes $\omega$.}. Occasionally, these are interpreted as transmission and reflection {\em probabilities}~\cite{hod1,ri_saa_1,hod2,matsasEtal} for a particle that is sent towards the black hole from infinity. Such an interpretation is not tenable, because for superradiant frequencies, the "transmission coefficient" becomes negative\footnote{In fact, \cite{hod1,hod2} use the phrase "negative probability".}.

The confusion seems to stem from the fact that despite the apparent analogy with the Schr\"{o}dinger equation, which is suitable for describing single particles, we deal here with relativistic (Klein-Gordon or Dirac) equations, which do {\em not} describe single particles, they {\em must} allow particle creation/destruction.  

If we consider the scattering of {\em classical} waves, the meanings of the coefficients are clear; they do represent ratios of energies coming back from the black hole and going into it, respectively, as can be verified by writing down the integrals for energy fluxes at infinity and the horizon, respectively, by using the stress-energy-momentum tensor of the field. The sum of the coefficients  is unity by virtue of eq.(\ref{eq:wronski}); a  manifestation of conservation of energy due to the stationary nature of the spacetime. For superradiant frequencies, the negativity of the transmission coefficient means that wave carries energy {\em out} of the black hole, hence the black hole's mass decreases, and the wave is seen coming back amplified by an observer at infinity. 

If we consider {\em quantum} waves, $AA^{*}$, the "reflection coefficient", represents the {\em expected value} of the ratio of the fluxes coming back from the black hole and going into it, respectively, at infinity. Since the wave mode has a given frequency, hence particles have a given energy (can be interpreted so at infinity), this coefficient is  proportional to the outgoing particle current. The fact that it can be larger than unity (i.e. superradiance) is a tip-off that particle creation is occuring, i.e. particle number is not conserved. Energy {\em is} conserved, however, so that black hole picks up the balance, even if it is negative. In this case presumably, a majority of particles going in through the horizon will have negative energies, which is possible inside the ergosphere. After all, superradiance only occurs for rotating black holes.

Because the reflection coefficient $AA^{*}$ represents the expected value of the relative ratio of the flux coming back, it can be written as
\begin{equation}
R = AA^{*} = \frac{1}{n_{i}}\sum_{n=0}^{\infty}nP(n) \label{ref_c_Ps}
\end{equation}
where $n_{i}$ is the number of incident particles, and $P(n)$ is the probability that $n$ particles will come back. Here we have assumed that the states representing different numbers of outgoing particles are orthogonal, and suppressed the dependence of the probabilities on $n_{i}$. The terminology "reflection" is misleading in this context, because even if we send in one particle and get one back, we do not know if the particle we catch is the same one that we sent in, or if that one entered the black hole and we caught the outgoing member of a produced pair. This terminology also conditions the mind into thinking in terms of particle number conservation. 

On the other hand, if we consider a {\em single quantum particle}, or a few ones, the $P(n)$'s are relevant instead of $A$. In other words, the reflection coefficient $AA^{*}$ {\em cannot} determine if a single particle will be absorbed or not; it cannot even determine the probability for this. Note that $A$ and $B$ can be found by solving eq.(\ref{eq:radial.u}) exactly. One can even find for which frequency range where we have superradiance, if any, by only solving the equation in the asymptotic regions; but the $P(n)$ cannot be found by solution of eq.(\ref{eq:radial.u}). For this, one needs to do a QFT calculation (e.g.~\cite{qft_capture_prob} or \cite{kermions}).

%%%%%%%%%%%%%%%%%%%%

\subsection{Quantum particles and WCCC}

Now, let us consider the case of small numbers of particles (per mode) more closely. Of course, we have
\begin{equation}
\sum_{n=0}^{\infty}P(n) = 1,   \label{Ps_normalztn}
\end{equation}
since {\em some} number of particles must come out, including possibly zero. Comparing eqs.(\ref{ref_c_Ps}) and (\ref{Ps_normalztn}), and considering $n_{i}=1$, we can see even without the QFT calculation  that when $R < 1$, e.g. for fermions, $P(0)$ must be positive. By continuity, we can expect $P(0)$ to be positive also for part of the $R>1$ (i.e. superradiant) range, if it exists. 

Without showing that $P(0)=0$, it cannot be claimed, as~\cite{hod1} does, that $\omega < \omega_{\rm sl}$ will mean that a scalar particle will not be absorbed. Therefore, the conclusion  of that work, that backreaction-induced\footnote{Incidentally, the backreaction argument of~\cite{hod1} is slightly puzzling: Instead of calculating in the {\em initial}, nonrotating spacetime, we are invited to calculate in the {\em final} spacetime, when the black hole has acquired the full angular momentum of the incoming particle. However, obviously both the initial and final spacetimes are equally representative or unrepresentative of the process, therefore it is hard to see why we should prefer one over the other. It sounds reasonable that calculating in the {\em average} spacetime would be a better way of taking backreaction into account. In fact, the paper's argument  mathematically works out also for $\omega = m^{2}/2M^{3}$, although in the main text we argued that the argument is fundamentally flawed.} superradiance will save WCCC in the thought experiment of~\cite{matsas_daSilva} is incorrect as well, in some sense making e.g.~\cite{matsasEtal} unnecessary. 

The fallacious notion of reflection/transmission coefficients as the respective probabilities, hence the incorrect claim of nonabsorption of scalar particles with energy/frequency in the superradiant range is accepted or propagated (at least partially) in~\cite{ri_saa_1,hod2,matsasEtal,ri_saa_2}. The paper~\cite{hod2} is the first one in the string to mention particle creation and absorption probability of single particles, yet also mentions transmission probability derived from solution of the field equation. Ref. \cite{matsasEtal} discusses behavior of an ensemble of particles vs. a single particle, yet still mentions the same probability. Even~\cite{ri_saa_2}, which clearly states that "superradiant modes have a nonzero probability of being absorbed", states also the opposite earlier in the paper. It seems that the multiparticle nature of the relativistic field equations and the limitations of the Schr\"{o}dinger analogy have not been completely appreciated, although hints exist.

So what {\em is} the problem with the Schr\"{o}dinger analogy? Should same equations not have same solutions? The problem is, eq.(\ref{eq:radial.u}) is {\em not} an eigenvalue problem for the energy ($\omega$), like a standard Schr\"{o}dinger problem; the potential depends on it.

%%%%%%%%%%%%%%%%%%%%

\subsection{Intermediate conclusion: Possible WCCC violation}

When the confusion about the absorption probabilities is cleared, the superradiance objections to the original Matsas-da Silva thought experiment~\cite{matsas_daSilva} evaporate; and the fermionic analog~\cite{ri_saa_1} is equally valid.  Therefore, sending tailored particles or waves into slightly subextremal black holes would seem to violate cosmic censorship, if backreaction and self-force effects are neglected. For single or few particles, quantum tunneling effects aid even when no violation would result for corresponding waves or classical test particles. Although self-force effects or a full treatment of backreaction could change the results, claims that backreaction via induced superradiance will prevent the violation appear to be not valid.

However, the claims that "vacuum polarization" will damp the absorption of tailored neutrinos by the black hole so that the best WCCC violation efforts will be undone by the effect, deserves closer scrutiny, which we undertake in the next section.

\section{The Zel'dovich-Unruh Effect as a Cosmic Censor} \label{s:Z-U}

The creation of particles discussed above is the analog of stimulated emission familiar from some other contexts, e.g. lasers. However, it turns out that an analog of {\it spontaneous} emission, which we will call the Zel'dovich-Unruh  effect\footnote{There seems to be no consensus in the literature on what to call this phenomenon. We think it is too specific to be called ``vacuum polarization''. Sometimes it is also called ``quantum superradiance'', sometimes it is associated with one or more of the names of Zel'dovich, Starobisnky (Starobinskii), and Unruh. Given the well-known Unruh effect, one cannot call it that. Some works, including ref.~\cite{kermions} call it the Unruh-Starobinskii effect, but the Starobinskii papers refer for this prediction to Zel'dovich's, which are earlier. Unruh brings a second quantization argument.}, also exists~\cite{zeldovich,zeldovich2,unruh}. 

This phenomenon emerges when one performs second quantization in the stationary, --i.e. eternal-- Kerr spacetime~\cite{unruh}. The rates at which the black hole loses mass and angular momentum due to this effect, i.e. the relevant fluxes at infinity, are given~\cite{dewitt} by
\begin{eqnarray}
& &\dfrac{\rm{d}M}{\rm{d}t}=\lim_{r\to \infty}\int d\theta d\phi \langle T_{rt}\rangle_{\rm{vac}} \sim -\frac{e^{-\zeta}}{4\pi}\Omega^2 \label{massflux} \\
& &\dfrac{\rm{d}J}{\rm{d}t}=-\lim_{r\to \infty}\int d\theta d\phi \langle T_{r\phi}\rangle_{\rm{vac}} \sim -\frac{e^{-\zeta}}{2\pi}\Omega \label{angmomflux}
\end{eqnarray}
where $\zeta$ is a number of the order of unity, and absolute units are used, i.e. $G=c=\hbar=1$.
It can be seen now that a higher fraction of angular momentum is emitted than mass/energy; more precisely,
\begin{equation}
\delta\left( \frac{a}{M} \right) \sim \frac{e^{-\zeta}}{4\pi} \frac{J}{2 M^{4} r_{+}}   \left( \frac{J^{2}}{M^{3} r_{+}} - 2 \right) 
 \label{ZU-CCC}
\end{equation}
is negative for all possible black hole parameters. So the Zel'dovich-Unruh effect always works towards preserving (the weak) cosmic censorship.

The question we ask at this point is if the WCCC-preserving effect of this quantum radiation is strong enough to invalidate the thought experiments \cite{matsas_daSilva,ri_saa_1,ri_saa_2} discussed above.

\subsection{The Zel'dovich-Unruh Effect and quantum tunneling arguments}

For the quantum tunneling thought experiments mentioned above, nearly extremal Kerr black holes are relevant, for which $J\sim M^2$, and (\ref{massflux}) and (\ref{angmomflux}) become 
\begin{equation}
\Delta J\sim -\frac{e^{-\zeta}}{4\pi} M^{-1}\Delta t \quad \mbox{;} \quad \Delta M\sim -\frac{e^{-\zeta}}{16\pi} M^{-2}\Delta t
\label{deltaj}
\end{equation}

Strictly speaking, (\ref{massflux}), (\ref{angmomflux}), and (\ref{deltaj}) give the the loss of mass and angular momentum due to emission of scalar field only; neutrinos are produced at a similar rate \cite{unruh} and photons and gravitons are produced more copiously \cite{staro2}. Therefore the rate of emission is about two orders of magnitude higher than these values \cite{dewitt}.

In the units adopted, even a (nearly extremal) super-massive black hole of $10^7$ solar masses has an angular momentum emission of the order of unity in a single second ($\Delta t=2 \times 10^{44}$ in absolute units). That is 20 orders of magnitude higher than the mass of the proton. The effect becomes stronger with decreasing mass, for example, the emission rate for a nearly extremal black hole of solar mass $\Delta J \sim 2\times 10^7$ in a second. This corresponds to a mass energy of $\sim 100$ g.
Therefore the Zel'dovich-Unruh effect makes thought experiments involving the tunneling of single or few particles with the purpose of WCCC violation completely meaningless.

Similar arguments apply to gedanken-efforts to overcharge nearly extremal Reissner-Nordstrom black holes. As studied by Gibbons \cite{gibbons},  charged black holes $(M,Q)$ emit particles (mass $m$, charge $e$) in the electrical superradiant region $\omega<eQ/r_+$ to neutralize themselves.  The charge loss is given by 
\begin{equation}
\frac{dQ}{dt} \sim \frac{1}{\exp [\frac{2\pi}{ \kappa} (\omega -eQ/r_+)]- 1} 
\label{dQdt1}
\end{equation}
For large black holes $(M\gtrsim10^{15}\rm{g})$ the rate of charge loss agrees with Schwinger's formula \cite{schwinger} for the rate of particles created by a uniform electric field,
\begin{equation}
\frac{dQ}{dt}\sim \frac{e^4 Q^3}{r_+}\exp \left( -\pi \frac{m^2}{e^2} \frac{r_+^2}{Q} \right) \label{discharge}
\end{equation}
where $m$ and $e$ are mass and charge of the electron. For a nearly extremal black hole, $r_+^2/Q \sim M$. Using $m^2/e^2 \sim 10^{-42}$ the argument of the exponential in (\ref{discharge}) is of the order of unity for back hole of $\sim 10^3 M_{\rm{s}}$. That leads to a charge loss proportional to $M^2$ which is much faster than angular momentum loss. 

For smaller black holes the charge flux is analogous to thermal process described by Hawking~\cite{hawking_rad}. In both regimes black holes rapidly discharge themselves, the process acting as a cosmic censor and completely  invalidating the efforts to over-charge them by absorption of a single or a few particles.

\subsection{Zel'dovich-Unruh Effect for scalar fields, without second quantization}

Since single- or few-particle thought experiments are shown to be irrelevant for WCCC violation studies, we have turn to the case of many, many particles. But then, we have to think in terms of ensembles and expected values, i.e. first-quantized fields, which in many respects should give the same answers as classical fields, which we have studied~\cite{dkn,overspin,koraydirac}. However, the Zel'dovich-Unruh effect, or at least a spontaneous emission, is not apparent in the usually used formalism: For example, the change in energy of a dyonic Kerr-Newman black hole as a result of interaction with a test scalar field is given as~\cite{dkn,overspin}
\begin{equation}
\delta M =  \frac{1}{2} \int d\omega \sum_{l,m} f_{lm}(\omega) f^{*}_{lm}(\omega)  \bar{\omega} \, \omega \, B_{l m}(\omega) B^{*}_{l m}(\omega) \label{scalardeltaM} 
\end{equation}
if one uses the standard normalization (\ref{solution_at_infty}); and where $f_{lm}(\omega)$ is the coefficient showing a mode's contribution to the wave packet. Spontaneous emission is the case of no incoming wave, which here can only be realized by setting all $f_{lm}(\omega)$ to zero; in which case $\delta M$ vanishes!

The problem is that  the standard normalization (\ref{solution_at_infty}), by setting the coefficient of the incoming part of the wave to one, {\em hides the possibility of the incoming component of a mode to vanish while keeping other components nonzero.} In other words, the normalization divides the wave by the amplitude of the incoming part before multiplying it with the (thought-) experimenter-configurable $f_{lm}(\omega)$; and the first step is a division by zero when there is no incoming wave.

Let us instead leave the amplitude free:
\begin{equation}
\lim_{r \rightarrow \infty} U_{l m}(\omega;r_{*})
                          = I_{lm}(\omega) e^{-i k r_{*}} + A_{lm}(\omega) e^{i k r_{*}},    	\label{solution_at_infty_new}
\end{equation}
where we think of the $I_{lm}(\omega)$ as user-configurable now. Then (\ref{eq:wronski}) becomes
\begin{equation}
\bar{\omega} B B^{*}  = k (I^{*} I - A^{*} A).\label{eq:wronski_new}
\end{equation}
where for given $I$ and $\omega$, in principle the differential equation (\ref{eq:radial.u}) can be solved to find one of $A$ and $B$, an then via (\ref{eq:wronski_new}) the other can be found. With this convention, the change in energy is written as
\begin{equation}
\delta M =  \frac{1}{2} \int d\omega \sum_{l,m} \bar{\omega} \, \omega \, B_{l m}(\omega) B^{*}_{l m}(\omega). \label{scalardeltaM_new} 
\end{equation}

We can now contemplate doing nothing to the black hole, i.e. having all $I=0$. Then in the range where both $\bar{\omega}$ and $k$ are positive (the non-superradiant range), both $A$ and $B$ have to vanish: nothing happens. But in the superradiant range, $\bar{\omega}$ is negative, and $\bar{\omega}$ and $k$ determine the ratio $|A/B|$. The expression (\ref{scalardeltaM_new}) is obviously negative when all $I_{l m}(\omega)$ are zero, although we cannot determine all $B_{l m}(\omega)$ from eq.(\ref{eq:wronski_new}).

This situation represents spontaneous creation of scalar particles, flowing both to infinity and into the black hole; a faraway observer will see the black hole as emitting particles. But, this first-quantized/semiclassical argument only gives the ratio of amplitudes flowing into and out of the black hole, it does not give the magnitude of the flux observed at infinity.

Therefore it can be said that the Zel'dovich-Unruh effect for scalars can actually be understood without second quantization. The fact that it occurs also for scalars suggest that the Zel'dovich-Unruh effect by itself is not related to the neutrino having only one helicity, as claimed in~\cite{hod2}, it is a more general phenomenon. However, this (semi)classical version is an hitherto neglected aspect of superradiance, therefore understanding spontaneous emission of fermions may in fact require second quantization.

\subsection{WCCC violation via waves?}

Coming back to the question of WCCC violation by sending many particles into black holes, we argue that the Zel'dovich-Unruh effect is automatically taken into account, at least semiclassically, when one changes the normalization convention. Also notice that the expression (\ref{scalardeltaM_new}) is simpler than the corresponding expression (\ref{scalardeltaM}). Similarly, the change in the electric charge is
\begin{equation}
\delta Q_{e} = \frac{e}{2} \int d\omega \sum_{l,m} \bar{\omega} \, B_{l m}(\omega) B^{*}_{l m}(\omega) \label{scalardeltaQ_new} 
\end{equation}
with the sign convention for four-current chosen such that $e$ is positive and positive frequency wave modes have positive charge density; and the change in angular momentum is 
\begin{equation}
\delta J =  \frac{1}{2} \int d\omega \sum_{l,m} \bar{\omega} \, m \,  B_{l m}(\omega) B^{*}_{l m}(\omega). 
\label{scalardeltaJ}
\end{equation}
Both integrands are negative in the superradiant region. $\delta Q_{e}$ and $\delta J$ are also negative when no incoming wave is present, i.e. all $I_{lm}(\omega)$ are zero. The first is the (semi)classical version of the spontaneous discharge discussed by Gibbons, mentioned above. While it may seem that there is no $Q_{e}$ dependence in $\delta Q_{e}$ [cf. eqs.(\ref{dQdt1}) and (\ref{discharge})], one should remember that the $B_{l m}(\omega)$ will depend on $Q_{e}$, moreover, for vanishing $I_{l m}(\omega)$, the $B_{l m}(\omega)$ will only be nonzero in the superradiant range, i.e. the limit of integration will be given in terms of $Q_{e}$.

Comparing the last three expressions to the corresponding ones in ref.~\cite{overspin}, one can see that the  argument of that work about WCCC violation by sending scalar waves onto a nearly extremal (dyonic) Kerr-Newman black hole carries through; the Zel'dovich-Unruh effect, at least in its (semi)classical form, does not change the argument, because it was already included.

%%%%%%%%%%%%%%%%%%%%%%%%%
\section{Conclusions}

In this work, we discuss several thought experiments tring to violate WCCC, involving quantum tunneling of single particles; first bosons, then fermions, carrying large angular momentum, into slightly subextremal black holes; and objections to them, namely that superradiance, or that failing for fermions, ``vacuum polarization'' or ``spontaneous emission of particles'' will uphold the WCCC.

We conclude that the superradiance objections are not valid, because the concepts of reflection/transmission coefficients do not represent probabililities for {\em single particles} to be reflected/absorbed; for the simple reason that the relevant equations are relativistic, hence allow particle creation.

However, the spontaneous emission objections are valid: This phenomenon, which we call the ``Zeldovich-Unruh effect'', completely dominates any single particles that may be sent into the black holes, and furthermore, acts as a cosmic censor. We also find that at least for scalars, the effect can be understood (semi)classically, i.e. without second quantization; this was hidden by the standard normalization convention.

If one stays in this framework, that is (semi)classical scalar fields, which should be equivalent to sending many quantum scalar particles, the conclusions of our previous work~\cite{overspin} stay valid, since the Zeldovich-Unruh effect had been included all along: WCCC may be violated by starting from a slightly subextremal black hole, and sending in (many) tailored particles.

The spontaneous emission of fermions, however, cannot be predicted (semi)classically. On the other hand, arguments that this emission will, by the exclusion principle, prevent absorption of incoming particles is not convincing, since it is not clear that the emission fills the phase space, especially considering that the incoming and outgoing momenta are oppositely directed. Therefore it would seem that sending fermion fields (many fermions) into a black hole~\cite{koraydirac} the Zeldovich-Unruh effect can be beaten (at least neutralized) and  WCCC violated. However, the Pauli principle may bring an upper limit to the fermion flux that can be sent in; the consideration of fermions will have to await further work.

%%%%%%%%%%%%%%%
\section*{Acknowledgements}

This work has been partially supported by the Bo\u{g}azi\,{c}i University Research Fund by grant number 14B03P3 (7981).

\end{document}